\documentclass[sigconf]{acmart}
\usepackage{booktabs} 
\usepackage{multirow}

\usepackage{color}
\usepackage{subfigure} 
\usepackage{array}
\usepackage{breqn}
\usepackage{{inputenc}}
\usepackage{balance}
\usepackage{multirow,tabularx}
\usepackage{longtable}
\usepackage{booktabs,longtable}
\usepackage{hhline}
\usepackage[flushleft]{threeparttable}
\usepackage{algorithm}
\usepackage{algorithmic}
\usepackage{epsfig}
\usepackage{graphicx}
\usepackage{epstopdf}
\usepackage{thmtools}
\usepackage{enumitem}
\usepackage{verbatim}
\usepackage{amsfonts}
\usepackage{hyperref}
\hypersetup{colorlinks=false,linkcolor=blue,urlcolor=blue,citecolor=red}
\usepackage{multirow}

\newcommand{\Lapl}{\mathbf{\mathop{\mathcal{L}}}}
\newcommand{\Trans}[1]{{#1}^{\top}}

\newcommand{\Mat}[1]{\mathbf{#1}}

\newcommand{\Space}[1]{\mathbb{#1}}
\newcommand{\Set}[1]{\mathcal{#1}}

\newcommand{\ie}{\emph{i.e., }}
\newcommand{\eg}{\emph{e.g., }}
\newcommand{\etal}{\emph{et al.}}

\newcommand{\wrt}{\emph{w.r.t. }}
\newcommand{\cf}{\emph{cf. }}
\newcommand{\aka}{\emph{aka. }}

\AtBeginDocument{%
  \providecommand\BibTeX{{%
    \normalfont B\kern-0.5em{\scshape i\kern-0.25em b}\kern-0.8em\TeX}}}

\copyrightyear{2020}
\acmYear{2020}
\setcopyright{acmcopyright}
\acmConference[SIGIR '20] {43rd International ACM SIGIR Conference on Research and Development in Information Retrieval}{July 25--30, 2020}{Virtual Event, China}
\acmBooktitle{43rd International ACM SIGIR Conference on Research and Development in Information Retrieval (SIGIR '20), July 25--30, 2020, Virtual Event, China}
\acmPrice{15.00}
\acmDOI{10.1145/3397271.3401080}
\acmISBN{978-1-4503-8016-4/20/07}

\begin{document}
\fancyhead{}

\title{
Hierarchical Fashion Graph Network for Personalized Outfit Recommendation
}


\author{Xingchen Li}
\affiliation{%
  \institution{Zhejiang University}
  }
\email{xingchenl@zju.edu.cn}

\author{Xiang Wang}
\affiliation{%
  \institution{National University of Singapore}
  }
\email{xiangwang@u.nus.edu}

\author{Xiangnan He}
\affiliation{%
  \institution{University of Science and Technology of China}
  }
\email{xiangnanhe@gmail.com}

\author{Long Chen}
\affiliation{%
  \institution{Zhejiang University}
}
\email{longc@zju.edu.cn}

\author{Jun Xiao}
\authornote{Jun Xiao is the corresponding author.}
\affiliation{%
  \institution{Zhejiang University}
  }
\email{junx@zju.edu.cn}

\author{Tat-Seng Chua}
\affiliation{%
  \institution{National University of Singapore}
  }
\email{dcscts@nus.edu.sg}

\begin{abstract}
Fashion outfit recommendation has attracted increasing attentions from online shopping services and fashion communities.Distinct from other scenarios (\eg social networking or content sharing) which recommend a single item (\eg a friend or picture) to a user, outfit recommendation predicts user preference on a set of well-matched fashion items.Hence, performing high-quality personalized outfit recommendation should satisfy two requirements --- 1) the nice compatibility of fashion items and 2) the consistence with user preference. However, present works focus mainly on one of the requirements and only consider either user-outfit or outfit-item relationships, thereby easily leading to suboptimal representations and limiting the performance.

In this work, we unify two tasks, fashion compatibility modeling and personalized outfit recommendation. Towards this end, we develop a new framework, \emph{Hierarchical Fashion Graph Network} (HFGN), to model relationships among users, items, and outfits simultaneously. In particular, we construct a hierarchical structure upon user-outfit interactions and outfit-item mappings. We then get inspirations from recent graph neural networks, and employ the embedding propagation on such hierarchical graph, so as to aggregate item information into an outfit representation, and then refine a user's representation via his/her historical outfits. Furthermore, we jointly train these two tasks to optimize these representations. To demonstrate the effectiveness of HFGN, we conduct extensive experiments on a benchmark dataset, and HFGN achieves significant improvements over the state-of-the-art compatibility matching models like NGNN~\cite{NGNN} and outfit recommenders like FHN~\cite{FHN}.
Our code has been released\footnote{\url{https://github.com/xcppy/hierarchical_fashion_graph_network}}.

\end{abstract}



\maketitle

\section{Introduction}

\begin{figure*}[t]
\centering
\includegraphics[width=1.9\columnwidth]{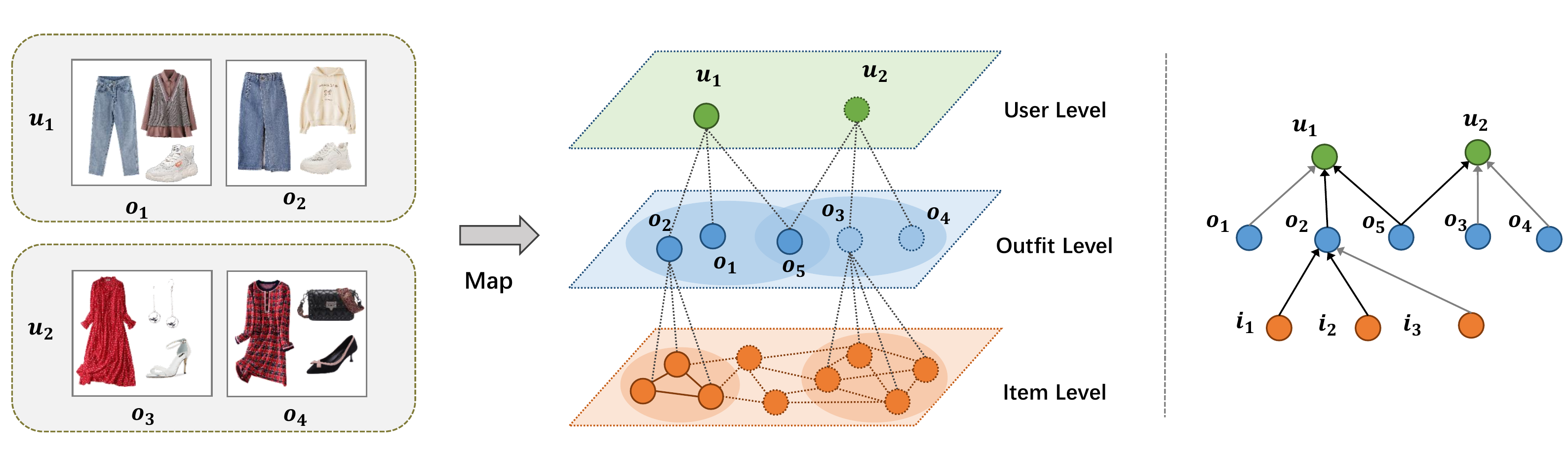}
\caption{The illustration of our hierarchical fashion graph network, HFGN. HFGN consists of three levels (\ie user, outfit and item level). The message can propagate from lower level to higher level. }
\vspace{-10px}
\label{fig:intro}
\end{figure*}

Fashion recommendation plays an activate role in daily life, with the rapid growth of online shopping platforms (\eg Amazon and Taobao) and fashion related social networks (\eg Instagram).
It can improve user experience and bring great profit to the shopping platforms.
Personalized outfit recommendation is an emerging service, which targets at selecting a set of visually-compatible items as an outfit for a target user.
Distinct from traditional fashion item recommendation which only routes a single product to users, it should satisfy two requirements --- 1) \emph{compatibility of fashion items}, meaning that the items within the same outfit should be visually compatible with each other, (\eg the long sleeve wear matches the high-heel shoes well in the outfit $o_{4}$ as Figure~\ref{fig:intro} shows); and 2)\emph{consistence with personal taste}, meaning that the whole outfit should holistically match user preference, \ie each user might have individual dressing style (\eg as Figure~\ref{fig:intro} shows, the outfit $o_{1}$ fits the casual style of user $u_{1}$, while user $u_{2}$ has special interests on the outfit $o_{4}$ due to its long sleeve style).

Nevertheless, most present works focus only on one of the requirements --- that is, either compatibility matching~\cite{NGNN} or outfit recommendation~\cite{FHN} --- while seldom modeling such two tasks simultaneously.
In particular, a research line on compatibility matching solely exploits the mapping relationships between outfits and single items to estimate whether multiple fashion items form a good match (\ie outfit).
For example, NGNN~\cite{NGNN} builds a fashion graph upon a taxonomy of fashion categories, representing an outfit as a graph instance involving compatible items as its nodes.
However, these works leave the personal interests of users untouched, hence typically serving as the tool to generate outfit features and failing to meet the second requirement of personalized recommendation.
Another straightforward solution to outfit recommendation is to subsume such task under the framework of traditional recommendation, relying largely on the user-outfit interactions.
Specially, the fashion items are simply seen as the pre-existing features of the outfit; as such, traditional recommenders can be adopted in such scenarios.
For example, VBPR~\cite{VBPR} and ACF~\cite{ACF} can be employed on users' historical interactions with outfits to perform future recommendation.
However, these methods are not tailored to outfit recommendation, forgoing the compatibility matching between fashion items.

Studies on jointly conducting compatibility modeling and outfit recommendation is under explored until recent FPITF~\cite{FPITF} and FHN~\cite{FHN}.
To be more specific, FPITF~\cite{FPITF} aggregates user preference on each item within an outfit, and integrates them with pairwise compatibility scores between items as a user's holistic preference on the outfit.
FHN~\cite{FHN} resorts to the similar paradigm.
Examining such paradigm, we find that the outfit-item mappings are treated as isolated data instance towards compatibility matching, while forgoing the relationships among data instances (\eg outfits with co-occurred items); analogously, the user-outfit interactions are fed individually into the recommender, while ignoring their relationships (\eg co-purchased outfits or behaviorally similar users).
This paradigm obscures the complex relationships among users, outfits, and items, easily leading to suboptimal representations and further limiting the recommendation performance.

In this work, to solve the foregoing limitations, we explicitly present the complex relationships among users, outfits, and items as a hierarchical fashion graph.
To be more specific, it consists of three levels --- user, outfit, and item levels --- where each level contains of the corresponding type of nodes.
Distinct from traditional graphs, such a hierarchical graph highlights the connections cross levels.
For example, if outfit $o$ contains item $i$, the node $o$ in the outfit level will be connected with the node $i$ in the item level;
analogously, if user $u$ has purchased outfit $o$, there is an edge between nodes $u$ and $o$ across the user and outfit levels.

Thereafter, we build a new framework, termed \textbf{Hierarchical Fashion Graph Network (HFGN)}, upon the hierarchical fashion graph.
In particular, HFGN employs the information propagation mechanism from graph neural networks (GNNs) to distill useful signals from the bottom to the top, inject the relationships into representations and facilitate the compatibility matching and outfit recommendation.
More specifically, we assign each user/outfit with an ID embedding while representing each item with its visual features.
The information propagation rule aggregates useful signals from fashion items to update outfit representations, and further refine user representations by integrating messages passing from his/her historical outfits.
Furthermore, we propose a joint learning scheme to conduct the compatibility matching and outfit recommendation simultaneously.
To demonstrate effectiveness of HFGN, we conduct extensive experiments, and the experimental results show that our model outperforms the state-of-the-art models \wrt both two tasks.

To summarize, we make the following main contributions in this paper:

\begin{itemize}
    \item We propose a Hierarchical Fashion Graph Neural Network (HFGN) to obtain more expressive representations for users and outfits. Benefiting from the message propagation rules, the representations can be updated by the neighbour embeddings iteratively.
    \item Different from the existing methods which only consider item-level semantics for outfits, we incorporate outfit-level semantics into the representations for outfits.
    \item Compared to separately considering compatibility matching and personalized recommendation, 
    we regard the compatibility information as a passing message in the graph and encode this information into item and outfit representations.
    Experiments show the rationality and effectiveness of this modeling.
    
\end{itemize}

\section{HFGN framework}
We now present our HFGN framework which is equipped with three key components: 1) embedding initialization, which initializes embeddings for user, outfit, and item nodes; 2) hierarchical graph convolution, which refines the node embeddings by propagating information from lower levels to higher levels --- that is, gather information from item nodes to update outfit representations, and then augment user representations via the historical outfits; and 3) model prediction, which outputs the prediction score for personalized recommendation and compatibility prediction.

\subsection{Embedding Initialization}

As Figure~\ref{fig:intro} shows, we organize users, outfits, and items in the form of a hierarchical fashion graph, where these three types of nodes are at the top, internal, and bottom levels, respectively.
To characterize the latent features, we represent each user/outfit/item ID with a vectorized representation (\aka embedding).
More formally, we denote IDs of user $u$, outfit $o$, and item $i$ separately as $\Mat{u}\in\Space{R}^{d}$, $\Mat{o}\in\Space{R}^{d}$, and $\Mat{i}\in\Space{R}^{d}$, where $d$ is the embedding size.

As a result, we can obtain an embedding table for all the nodes as follows:

\begin{equation}
    \Mat{E} = [\underbrace{\cdots,\Mat{i},\cdots}_{\rm item\:  embeddings},\underbrace{\cdots,\Mat{o},\cdots}_{\rm outfit\: embeddings},\underbrace{\cdots,\Mat{u},\cdots}_{\rm user\: embeddings}],
\end{equation}
where $\Mat{E}\in\Space{R}^{(N_{U}+N_{O}+N_{I})\times d}$ concatenates embeddings of users, outfits, and items; $N_{U}$, $N_{O}$, and $N_{I}$ are the number of users, outfits, and fashion items, respectively.

As only ID information is available for users and outfits, we get inspirations from the mainstream CF models~\cite{BPRMF,NCF} and project each user/outfit ID into an embedding. Such trainable embeddings are used to memorize the latent features of users and outfits.
As for each fashion item $i$, we have its visual feature $\Mat{x}_{i}$.
Moreover, since items are associated with different fashion categories (\eg jeans, T-shirt), we use category-aware encoders to extract useful information from the visual features as item embeddings.
More formally, the initial embeddings of items are encoded as:
\begin{gather}
    \Mat{e}_{(i)} =  f_{c}(\Mat{x}_{i}),
\end{gather}
where $f_{c}(\cdot)$ is the encoder for category $c$, which is implemented by a two-layer MLP.
As such, the item features are projected into the same latent space to that of users and outfits, facilitating the further modeling of their complex relationships.

\begin{figure*}[ht]
\centering
\subfigure[Information propagation across items.]{
\label{fig:itemUpdate}
\centering
\includegraphics[width=0.32\textwidth]{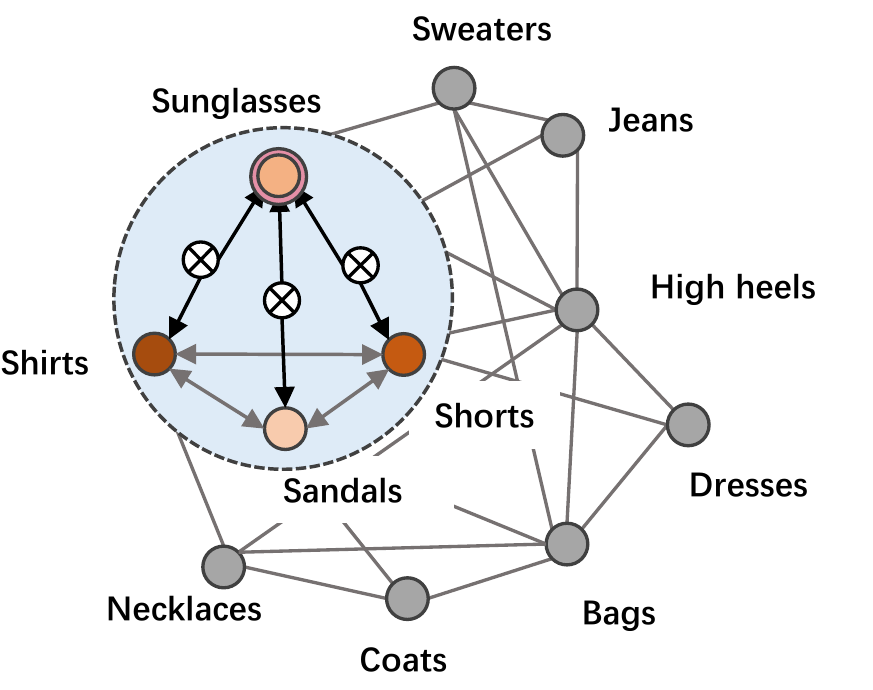}}
\subfigure[Information propagation from item to outfit level.]{
\label{fig:outfitUpdate}
\centering
\includegraphics[width=0.32\textwidth]{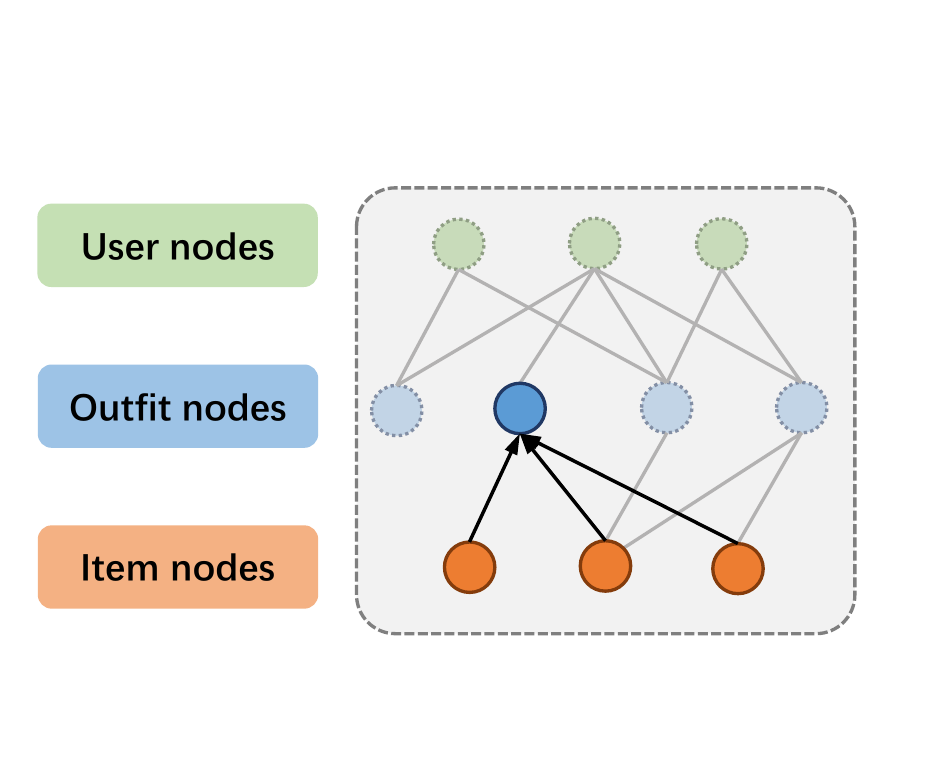}}
\subfigure[Information propagation from outfit to user level.]{
\label{fig:userUpdate}
\centering
\includegraphics[width=0.32\textwidth]{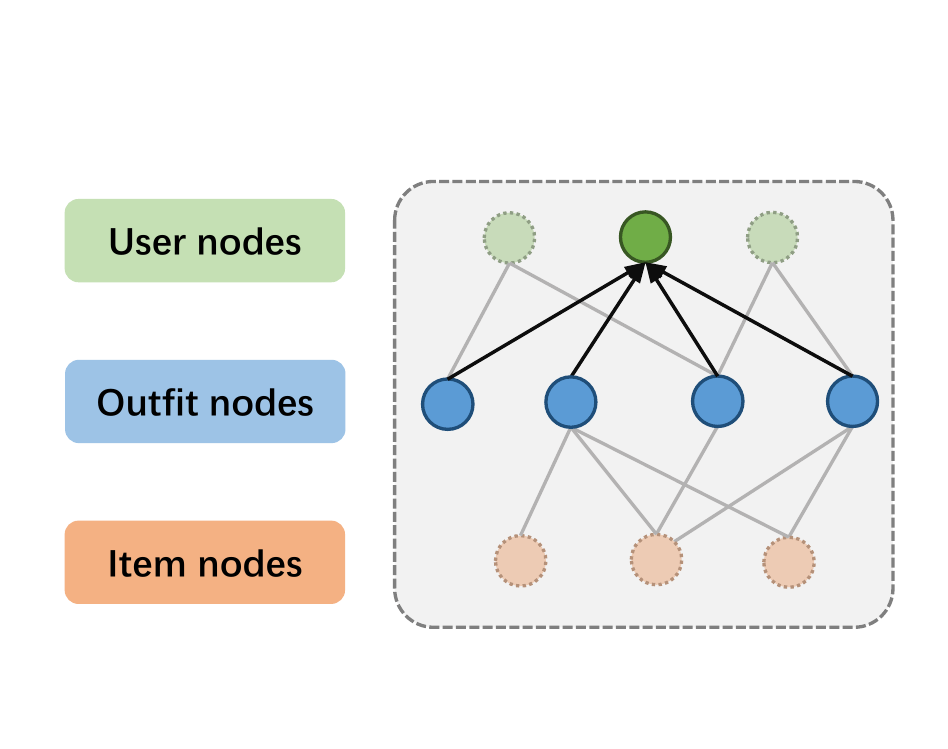}}
\caption{Information propagation across three levels in HFGN. (a) presents information propagation across items; (b) presents information propagation from item to outfit level; (c) presents information propagation from outfit to user level.}
\label{fig:embeddingUpdate}
\end{figure*}

\subsection{Hierarchical Graph Convolution}

By organizing users, outfits, and items in a hierarchical graph, we can leverage their connectivities to help exhibit their underlying relationships.
For example, As the right in Figure~\ref{fig:intro} shows, the path $i_{1}\rightarrow o_{2}$ states the fact that outfit $o_{2}$ includes item $i_{1}$, while $o_{2}\rightarrow u_{1}$ presents the behavior that user $u_{1}$ purchases outfit $o_{2}$; meanwhile, the longer path $i_{1}\rightarrow o_{2}\rightarrow u_{1}$ might reflect user $u_{1}$'s interest on item $i_{1}$, while $\{i_{1},i_{2}\}\rightarrow o_{2}$  and $ o_{5}\rightarrow \{u_{1}, u_{2}\} $ separately indicate the compatibility of items and behavioral similarity of users. 
Hence, exploiting such connectivities is of crucial importance to explore relationships among users, outfits, and items, and is a promising solution to unify compatibility modeling and outfit recommendation.

Recent studies on graph neural networks~\cite{GCN,GAT,graphsage} show that the information propagation over graph structure is able to effectively distill useful information from multi-hop neighbors and encode higher-order connectivity into the representations.
Inspired by this, we devise a new hierarchical graph convolution (HGC) to perform the embedding propagation mechanism over our fashion graph, so as to refine their embeddings.
In particular, there are three embedding-propagation steps --- 1) information propagation across items, which refines item embeddings by incorporating the compatibility modeling; 2) information propagation from items to outfits, which aggregates item semantics into outfit embeddings; and 3) information propagation from outfits to users, which integrates historical outfits as user representations.
In what follows, we elaborate these three ingredients.

\subsubsection{\textbf{Information Propagation Across Items}}
Items are at the lowest level of HFGN, providing visual features of individual items, as well as their compatible relationships.
For example, the connectivities $\{i_{1},i_{2}\}\rightarrow o_{2}$ not only describe that items $i_{1}$ and $i_{2}$ belong to the same outfit $o_{2}$, but also reflect that $i_{1}$ and $i_{2}$ are compatible.
Hence, such compatibility information suggests that compatible items should have more information interchange than that in different outfits.
Towards presenting compatibility among items in an explicit fashion, we construct an item graph for each outfit first.

\vspace{5pt}
\noindent\textbf{Item Graph Construction.}
Before constructing item graphs for individual outfits, we first build a uniform category graph~\cite{NGNN} for all outfits, where the category information of items serve as the prior knowledge of items.
In particular, each item is assigned with only one specific category, such as \emph{shirts}, \emph{sandals}, and \emph{jeans}.
Different category pairs are associated with varying co-occurrence frequencies, reflecting the coarse-grained compatibility of items at a granular level of category.
For instance, \emph{necklaces} co-occurs more frequently with \emph{coats} in outfits, than \emph{sandals}.
Hence, we build a weighted category graph $\Set{G}_{\text{c}}=\{(c,c',w(c,c'))|c,c'\in\Set{C}\}$, where $\Set{C}$ is the set consisting of 60 categories in total (\cf data statistics in Table~\ref{tab:fltb_dataset}).
Wherein, each category pair $(c,c')$ is assigned with a weight as follows:
\begin{gather}
    w(c,c') = \frac{g(c,c')/g(c')}{\sum_{c''\in\Set{C}}g(c,c'')/g(c'')},
\end{gather}
where $g(c,c')$ denotes the co-occurrence frequency of categories $c$ and $c'$ appearing in the same outfits, while $g(c)$ counts the frequency of $c$ in the outfit-item and item-category mappings.

Having established $\Set{G}_{\text{c}}$ for all outfits, we now construct an item graph tailored $\Set{G}_{\text{o}}$ for a single outfit $o$.
In particular, we activate the category nodes which appear in outfit $o$ (\eg the orange nodes shown in Figure~\ref{fig:itemUpdate}), while removing the others.
More formally, $\Set{G}_{\text{o}}$ is defined as $\{(c,c',w(c,c'))|c,c'\in \Set{N}_{o}\}$, where $\Set{N}_{o}$ is the item set of outfit $o$.
Clearly, the weights of $\Set{G}_{\text{o}}$ directly inherit from the original category graph, and only parts of nodes with their connections are activated as the blue circle in Figure~\ref{fig:itemUpdate} shows. 

\vspace{5pt}
\noindent\textbf{Item-Wise Information Construction.}
Presenting the coarser-grained compatibility in the form of item graph, we now focus on one specific item and distill useful signals from its neighbors, where the compatibility \wrt categories is encoded.
In particular, the information being propagated from the neighboring item $i'$ to the ego item $i$ is formalized as:
\begin{gather}
    \Mat{m}_{i'\rightarrow i}=w(i,i')\sigma(\Mat{W}_{1}(\Mat{i}\odot\Mat{i}')),
\end{gather}
where $\Mat{W}_{1}\in\Space{R}^{d\times d}$ is the trainable weight matrix to perform transformation; $\sigma(\cdot)$ is a nonlinear activation function, which is set as LeakyReLU~\cite{leakyrelu} here; $\odot$ is the element-wise product. Hence, the signal $\Mat{i}\odot\Mat{i}'$ accounts for the visual compatibility between items $i$ and $i'$, encouraging compatible items to contribute more.
Furthermore, the weight $w(i,i')$ takes the categorical compatibility into consideration to control how much signals are being passed across categories.
Such weights also can be seen as the discounting factors adopt in GNN models~\cite{GCN,GAT}.

\vspace{5pt}
\noindent\textbf{Cross-Item Information Aggregation.}
For each item node, we can leverage the signals, which are pertinent to its affinity with co-occurred items (\ie neighbors), to update its embedding.
Here we employ the sum aggregator on item $i$'s neighbors as follows:
\begin{gather}
    \Mat{i}^{*}=\Mat{i}+\sum_{i'\in\Set{N}_{i}}\Mat{m}_{i'\rightarrow i},
    \label{equ:item-representations}
\end{gather} 
where $\Mat{i}^{*}$ is the updated embedding of item $i$.
Here only the sum aggregator is applied, leaving the exploration of other aggregators, such as attention networks~\cite{GAT,SCA-CNN}, in future work.
As a result, the compatibility information carried in the first-order connectivity are encoded into the item embeddings.
We can stack more layers to synthesize richer semantics in higher-order connectivity, leaving this exploration in future work.

\subsubsection{\textbf{Information Propagation from Item to Outfit Level}}
Intuitively, an outfit can be described by its involved items.
Taking Figure~\ref{fig:intro} as an example, outfit $o_{1}$ consists of a sweater, jeans, and running shoes, while outfit $o_{2}$ is composed of a pullover, jeans, and sneakers.
Rich item features help reveal underlying relationships between outfits \wrt visual similarity and categorical compatibility.
For example, outfits $o_{1}$ can serve as a substitute for $o_{2}$, due to the closely compatibility style; meanwhile, the style of outfit $o_{4}$ is different from that of $o_{1}$ and $o_{2}$, since they have no overlapping categories or items.
We hence augment ID embeddings of outfits with the representations of the involved items, to improve the quality of embedding.
In particular, we build a heterogeneous graph involving item and outfit nodes, where the edges are the item-outfit links.

\vspace{5pt}
\noindent\textbf{Item-Wise Information Construction.}
Focusing on an outfit, we refine the information that are influential to it from the involved items.
Formally, the messages being passed from the neighboring item $i$ to the ego outfit $o$ is:
\begin{gather}
    \Mat{m}_{i\rightarrow o}=\frac{1}{|\Set{N}_{o}|}\sigma(\Mat{W}_{2}\Mat{i}^{*}),
\end{gather}
where $\Mat{W}_{2}\in\Space{R}^{d\times d}$ is a trainable matrix to perform transformation; $\frac{1}{|\Set{N}_{o}|}$ is the normalization term to handle the varying number of involved items and stabilize the training.

\vspace{5pt}
\noindent\textbf{Outfit-Wise Information Aggregation.}
Analogous to the cross-item information aggregation, we combine the information from all involved items together as the final representation of an outfit, as follows:
\begin{gather}
    \Mat{o}^{*}=\Mat{o}+\sum_{i\in\Set{N}_{o}}\Mat{m}_{i\rightarrow o}.
\end{gather}
Obviously, the refined representation $\Mat{o}^{*}$ is composed of the ID embeddings and item-aware features.
Different from prior studies~\cite{FPITF,FHN} that build outfit representations upon the visual features of items solely, this information aggregation additionally considers the compatibility scores.

\subsubsection{\textbf{Information Propagation from Outfit to User Level.}}
Present studies~\cite{FISM,NAIS,SVD++} have shown that personal history directly profiles a user's preference.
For instance, analyzing the historical outfits (\ie $o_{1}$ and $o_{2}$) of user $u_{1}$, we might summarize her dressing style; moreover, $o_{5}\rightarrow \{u_{1}, u_{2} \}$ indicates the behavioral similarity between user $u_{1}$ and $u_{2}$.
Furthermore, collected personal histories reflect CF signals~\cite{NGCF,Hoprec}, referring that behaviorally similar users would have similar preference on outfits.
We hence enrich ID embeddings of users by incorporating the representations of historical outfits.
Towards that, we organize the user-outfit interactions in the form of heterogeneous graph.

\vspace{5pt}
\noindent\textbf{Outfit-Wise Information Construction.}
For a target user $u$, we focus on the items he/she adopted before $\Set{N}_{u}$, and extract useful information from each outfit $o$ as follows:
\begin{gather}
    \Mat{m}_{o\rightarrow u}=\frac{1}{|\Set{N}_{u}|}\sigma(\Mat{W}_{3}\Mat{o}^{*}),
\end{gather}
where $\Mat{W}_{3}\in\Space{R}^{d\times d}$ is the transformation matrix; here we assume that different outfits might contribute equally to profile a user, hence using $\frac{1}{|\Set{N}_{o}|}$ as priors and leaving the exploration of attentive weights in future.

\vspace{5pt}
\noindent\textbf{User-Wise Information Aggregation}
Thereafter, we employ the sum aggregator on the historical items, updating the user's representation as follows:
\begin{gather}
    \Mat{u}^{*}=\Mat{u}+\sum_{o\in\Set{N}_{u}}\Mat{m}_{o\rightarrow u},
\end{gather}
where the final representation $\Mat{u}^{*}$ consists of two components --- the ID embedding, which characterizes the intrinsic features of $u$, and the outfit-aware features, which present her dressing style explicitly.
Distinct from prior works~\cite{FPITF,FHN} which use ID embeddings for users only, our HFGN results in better representation ability.

After propagating information within the hierarchical fashion graph, we allow the information flow from the bottom to the top levels, exploiting the complex relationships among items, outfits, and users to guide the representation learning.

\subsection{Model Prediction}

Thereafter, we propose a joint learning scheme to conduct the compatibility matching and outfit recommendation simultaneously.

\subsubsection{\textbf{Personalized Outfit Recommendation}}
To predict how likely user $u$ would purchase outfit $o$, we employ the inner product on their representations as:
\begin{gather}
    \hat{y}_{uo}=\Trans{\Mat{u}^{*}}\Mat{o}^{*},
\end{gather}
which casts the predictive task as the similarity estimation between $u$ and $o$ in the same latent space.
As the main focus of this work is representation learning, we leave the exploration of interaction modeling in future work.

\subsubsection{\textbf{Compatibility Matching}}
To estimate whether multiple fashion items form a good outfit, we utilize the item representations (\cf Equation~\ref{equ:item-representations}) to calculate the matching score.
Distinguished from present works~\cite{FHN,FPITF} that simply aggregate the pairwise compatibility scores together, we argue that items have varying importance for the outfit.
For example, in outfit $o_3$, as the long dress determines the holistic style, it is more important than the accessories.
We hence differentiate importance of items in one outfit via a self-attention mechanism, which generates a $R$-view attention map and a $R$-view score map.
Formally, the attention map is calculated as:
\begin{gather}
    \Mat{A}=\rho\Big(\Mat{W}_{4}\sigma(\Mat{W}_{5}\Trans{\Mat{I}})\Big),
\end{gather}
where $\Mat{I}\in\Space{R}^{ n\times d}$ is embedding matrix of one outfit and $n$ is the length of the outfit;
 $\Mat{W}_{4}\in\Space{R}^{R\times v}$ and $\Mat{W}_{5}\in\Space{R}^{v\times d}$ are two trainable weight matrices.
 After transformation by $\Mat{W}_{4}$ and $\Mat{W}_{5}$, we can get an $R$ views attention map $\Mat{A}\in\Space{R}^{r\times n}$.
 $\rho(\cdot)$ is set as the softmax function to normalize the attention scores over items.
 
Thereafter, we establish the $R$-view score map for each outfit, considering each view as a latent factor which is influential for compatibility, as follows:
\begin{gather}
    \Mat{C}=\sigma\Big(\Mat{W}_{6}\sigma(\Mat{W}_{7}\Trans{\Mat{I}})\Big),
\end{gather}
where $\Mat{W}_{6}\in\Space{R}^{R\times v}$ and $\Mat{W}_{7}\in\Space{R}^{v\times d}$ are weight matrices.
As such, we project the original item representations into a latent space, describing each item from multiple factors.

Based on such attention and score maps, we get the weighted compatibility score of the outfit $o$ as follows:
\begin{gather}
    \hat{s}_{o}=\sum_{r=1}^{R}\Trans{\Mat{a}}_{r}\Mat{c}_{r},
\end{gather}
where $\Mat{a}_{r}\in\Space{R}^{n}$ and $\Mat{c}_{r}\in\Space{R}^{n}$ is the $r$-th rows of $\Mat{A}$ and $\Mat{C}$, respectively.

\subsection{Optimization}
In the following, we introduce the objective function for our model and training strategy.
\subsubsection{\textbf{Objective function}}
We adopt Bayesian Personalized Ranking (BPR) algorithm \cite{BPRMF} for both tasks. BPR assumes the observed interaction has higher prediction scores than unobserved ones.
The objective functions for two tasks are,
\begin{equation}
   \Lapl_{mf} = \min_{\Theta}\sum_{(u,o,o')\in\Set{H}}-\ln{\mu(\hat{y}_{uo}-\hat{y}_{uo'})},
\end{equation}

\begin{equation}
   \Lapl_{com} = \min_{\Theta}\sum_{(o,o')\in\Set{H}'}-\ln{\mu(\hat{s}_{o}-\hat{s}_{o'})},
 \end{equation}
where $\Set{H}=\{(u,o,o')\}$ is the training set for outfit recommendation, where each triple $(u,o,o')$ denotes user $u$'s historical interaction with outfit $o$ and a unobserved interaction with outfit $o'$; $\Set{H'}=\{(o,o')\}$ is the training set for compatibility modeling, where each pair $(o,o')$ denotes the observed outfit $o$ (\ie positive samples) and a unobserved outfit $o'$ (\ie negative samples by randomly generated);
$\mu(\cdot)$ is the sigmoid function;
and $\Theta$ is the set of model parameters, on which $L_{2}$ normalization is conducted to avoid overfitting.

\subsubsection{\textbf{Training Strategy}}
Due to the imbalance of the training data (\ie recommendation task has 1.6M positive samples while compatibility task has 0.027M positive samples in our dataset shown in table~\ref{tab:recom_dataset} and table~\ref{tab:fltb_dataset}.) and different model convergence speeds, we set different learning rates for two tasks.
And we jointly optimize $\Lapl_{mf}$ and $\Lapl_{com}$ in one epoch in our training process.

\section{Experiment}
We perform our experiments on a benchmark dataset, POG~\cite{POG}, and we aim to answer the following questions:

\begin{itemize}
    \item \textbf{RQ1}: How does HFGN perform as compared with the state-of-the-art methods on personalized outfit recommendation task?
    \item \textbf{RQ2}: How does the message propagation on each level affect HFGN?
    \item \textbf{RQ3}: How does HFGN perform as compared with the state-of-the-art methods on compatibility matching task?
\end{itemize}

\subsection{Experiment settings}
\subsubsection{\textbf{Dataset}}
As personalized outfit recommendation task is a fresh task in fashion community, the avaliable dataset is limited. POG~\cite{POG} is the only available large-scale dataset that meets our requirements.
The dataset collects click actions from 3.57 million users, including 1.01 milion outfits and 583 thousand individual items with context information.

POG dataset provides three files: 
1) user data, which records the user clicks on both outfits and items,
2) outfit data, which lists the items composed of the outfit,
and 3) item data, which provides the context details of items including categories, text descriptions and image download links. 
In our experiment, we utilize the user clicks on outfits to construct our dataset. We extract visual features for items from a pretrained Resnet-152. 

\vspace{5pt}
\noindent\textbf{Dataset for recommendation.}
To ensure the quality of dataset, we only retain the users with at least 20 interactions and the outfits with at least 10 interactions. The number of item categories is $60$. The statistics of recommendation dataset is shown in Table~\ref{tab:recom_dataset}.
To evaluate the recommendation performance, we randomly select $80\%$ of click histories of each user to build the training set, and take the remaining histories as the test set. We also split $10\%$ of training set as validation set to tune the hyper-parameters.

\begin{table}
\centering
\caption{Statistics of the dataset for personalized outfit recommendation.}
\label{tab:recom_dataset}
\resizebox{0.70\columnwidth}{!}{\begin{tabular}{c|c|c|c}
\hline 
 $\#$Users & $\#$Outfits & $\#$Interactions & $\#$Items \\
\hline\hline
  53,897 & 27,694    & 1,679,708      & 42,563 \\
\hline
\end{tabular}}
\end{table}

\vspace{5pt}
\noindent\textbf{Dataset for compatibility matching.}
We take the positive outfits in recommendation dataset as training set for compatibility matching, and additionally select $6,924$ ($10\%$ of training dataset) outfits that has never appeared in training set to construct test dataset. 
The statistics are shown in Table~\ref{tab:fltb_dataset}.
To evaluate the performance of HFGN on compatibility matching, we adopt a widely used task, Fill-in-the-Blank~\cite{bi-lstm}. 

\vspace{5pt}
\noindent\textbf{Fill-in-the-Blank (FLTB)}.
Fill-in-the-Blank (FLTB)~\cite{bi-lstm} is a task to select the most compatible item from the candidates to fill in the blank in one outfit, as Figure~\ref{fig:caseStudy} shows. 
For each outfit in test dataset, we randomly mask one item as a blank, and then randomly select 3 items from other outfits to form a candidate set with the masked item.
We set the masked item as the ground truth, assuming that the masked item is more compatible than other candidates.

\begin{table}
\centering
\caption{Statistics of the dataset for compatibility matching.}
\label{tab:fltb_dataset}
\resizebox{0.75 \columnwidth}{!}{\begin{tabular}{c|c|c|c}
\hline
 Dataset & \#Outfits & \#items  & \#categories\\
\hline\hline
Training set & 27,694  & 42,563 & 60\\
Test set & 6,924  & 15,984 & 60\\
\hline
\end{tabular}}
\end{table}

\subsubsection{\textbf{Parameter Settings}}
We perform grid search to tune the hyper-parameters for our model and baselines.
We search the batch size in $\{256, 512, 1024\}$,
and we tune regulation rate and learning rate in $\{10^{-5},10^{-4},10^{-3},10^{-2}\}$ and $\{0.0001,0.0005,0.001,0.005,0.01\}$, respectively.
Moreover, we optimize all the models with Adam optimizer.

\subsection{Personalized Outfit Recommendation }
In this section, we will discuss the performance of our model on personalized outfit recommendation task compared to the state-of-the-art methods.

\subsubsection{\textbf{Evaluation Metric}} 
To assess the performance of Top-$K$ recommendation, we adopt four widely-used metrics~\cite{NGCF,NCF}: HR@K, NDCG@K, Recall@K and Precision@K.
\begin{itemize}
    \item HR@$K$ measures weather any positive samples present in the top $K$ position. It is 1 if yes, otherwise 0.
    \item Recall@$K$ is the proportion of positive samples that have been successfully recommended to the user.
    \item Precision@$K$ is the proportion of the recommended items that are relevant to the user.
     \item NDCG@$K$ is a widely used measure to evaluate the quality of the ranked list, considering the graded relevance among positive ad negative samples within the top $K$ ranking list.
\end{itemize}
We set $K$ as $10$ in our experiment. For all the metrics, we report the mean of all users as the final score.

While the validation set is used to tune the hyper-parameters, we report the performance on the test set.

\begin{table}[t]
\centering
\caption{Overall performance comparison on personalized outfit recommendation.$*$ denotes the statistical significance for $p<0.001$.}
\label{tab:recommendationCompare}
\resizebox{0.95\columnwidth}{!}{
\begin{tabular}{l|c c c c}
\hline
      & HR@10          & NDCG@10        & Recall@10      & Precision@10 \\
\hline\hline
FPITF    & 0.1006       & 0.0420       & 0.0183          & 0.0112 \\
FHN    & 0.1109         & 0.0490        & 0.0208         & 0.0119 \\
\hline
MF     & 0.2121         & 0.0872        & 0.0434         & 0.0239 \\
VBPR   & 0.2201         & 0.0949        & 0.0449         & 0.0248 \\
\hline
NGCF   & $\Mat{0.2619}$ & $\Mat{0.1143}$ & $\Mat{0.0554}$ & $\Mat{0.0310}$\\ 
\hline
HFGN   & ${\Mat{0.2833}^{*}}$  & ${\Mat{0.1241}^{*}}$ &  ${\Mat{0.0605}^{*}}$  & ${\Mat{0.0339}^{*}}$ \\
\hline\hline
\%Improv. & 8.17\%  &  8.57\%  &  9.20\%  & 9.35\%\\
\hline
\end{tabular}}
\vspace{-10px}
\end{table}

\subsubsection{\textbf{Baselines}} 
As the personalized outfit recommendation task is fresh, and only a few works~\cite{FPITF,FHN} has researched on it. 
To demonstrate the effectiveness of our proposed model, we select some traditional recommendation models (\eg MF, VBPR) as our baselines. 
Besides, we also select a graph-based recommendation model to compare since we exploit graph structure in our method.

\begin{itemize}
    \item \textbf{FPITF}~\cite{FPITF}: FPITF represents users and items with ID embeddings and visual features respectively. The prediction score is composed of user's preference score for each item within one outfit and the compatibility score for each item pair in the outfit.
    \item \textbf{FHN}~\cite{FHN}: FHN encodes the visual features with category encoders and then learn binary code for both user and item embeddings. 
    The final score is also composed of two parts, preference score for user-item pairs and compatibility score for item-item pairs. For each part, FHN additionally introduces a diagonal weighting matrix to better model the interactions.
    \item \textbf{MF}~\cite{BPRMF}: Matrix Factorization (MF) model is one of the most popular techniques in personalized recommendation task. In this model, users and items are both projected into a same latent space and represented by the vectors in this space and their interaction score is calculated by inner product. 
     \item \textbf{VBPR}~\cite{VBPR}: Compared to MF model, VBPR additionally considers the user preference on visual factors. Here, we represent the visual feature of one outfit by averaging the visual features of the items within the outfit.
    \item \textbf{NGCF}~\cite{NGCF}: NGCF utilizes graph structure to model the high-order interaction between users and items. 
    Node embeddings are refined by stacking multiple propagation layers in interaction graph.
    Here, we conduct a two-layer propagation and the nodes in interaction graph are users, outfits and items.
    
\end{itemize}

\begin{figure}[t]
\centering
\includegraphics[width=0.98\columnwidth]{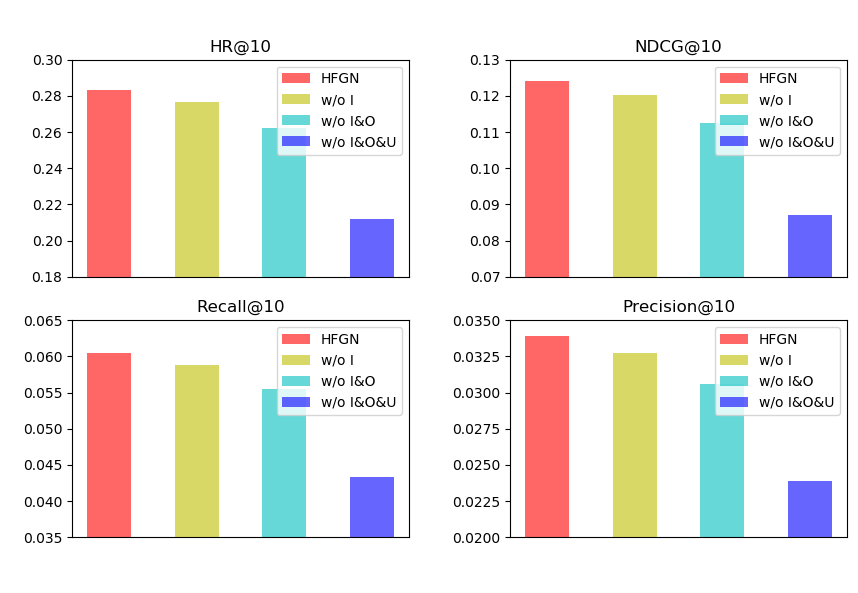}
\vspace{-5px}
\caption{Performance comparison of ablation study on HFGN.}
\vspace{-10px}
\label{fig:ablationStudy}
\end{figure}

\subsubsection{\textbf{Performance Comparsion (RQ1)}} 
Table \ref{tab:recommendationCompare} reports the performance results on personalized outfit recommendation. From the table, we have the following observations:

\begin{itemize}
    \item FPITF and FHN perform much worse than other baselines. 
    In both FPITF and FHN, they evaluate the user preference on one outfit only by averaging the preference score on each item. 
    However, outfits and items have different semantics, it's insufficient to only consider item-level preference.
    In other models, they employ ID embeddings of outfits to capture this information, enhancing the performance greatly in personalized recommendation task. 
    \item Compared to BPR, VBPR achieves better performance. Such improvement indicates the importance of incorporating visual signals into prediction formulations.
    \item From the Table~\ref{tab:recommendationCompare}, we can observe that NGCF performs much better than MF and VBPR. Benefiting from multiple propagation layers in interaction graph, NGCF is capable of modeling high-order connectivities among users, outfits and items.
    \item HFGN yields the best performance on all the metrics. 
    In particular, HFGN achieves the improvement over the strongest baseline (\ie NGCF) of $9.35\% $ \wrt Precision@10.
    Benefiting from message propagation mechanism, HFGN is capable of distilling useful signals from the bottom to the up and modeling complex relationships including user-outfit interactions and outfit-item mappings.
\end{itemize}

\subsubsection{\textbf{Study of HFGN (RQ2)}} 
To demonstrate the effectiveness of message propagation on each level, we conduct an ablation study. 
We disable the message propagation for each level and compare the performance, the results are shown in Figure~\ref{fig:ablationStudy}.

\begin{itemize}
    \item First, we disable the message propagation on item level, termed HFGN$_{\text{w/o I}}$. The message discarded in this operation is the compatibility of pair of items in one outfit, which is an important factor that influences user's interest. 
    From Figure~\ref{fig:ablationStudy}, we can see that HFGN$_{\text{w/o I}}$ slightly underperforms HFGN. 
    It seems that discarding this part of information doesn't hurt the performance much in our experiment.
    That's because the negative samples in our test dataset is the remaining outfits that users haven't clicked. These outfits are collected directly from the website and have a good compatibility.
    Therefore, the compatibility influences slightly on evaluation results.
    In fact, our model has a good capability to analysis compatibility, which will be discussed in Section~\ref{sec:compatibilityModeling}.
    As we mainly research on the architecture of personalized outfit recommendation,
    we leave the expansion of the test dataset in future work.
    \item Then, we disable the message propagation from item to outfit level, termed HFGN$_{\text{w/o I\&O}}$. 
    It means the representation of an outfit only has its ID embedding without item information. 
    From Figure~\ref{fig:ablationStudy}, we can see that HFGN$_{\text{w/o I\&O}}$ performs worse than both HFGN and HFGN$_{\text{w/o I}}$. 
    It makes sense since that item information plays a significant role in modeling outfit representation. 
    It hence illustrates the rationality and effectiveness of our message propagation rule from item to outfit level.
    \item Finally, we disable the message propagation from outfit level to user level, termed HFGN$_{ \text{w/o I\&O\&U}}$. It means that we only utilize ID embeddings for both users and outfits, which equals to MF model. 
    From Figure~\ref{fig:ablationStudy}, we can observe that the performance decreases greatly compared to the models mentioned above.
    Compared to HFGN$_{\text{w/o I\&O}}$, the message discarded in this operation is the historical interactions of users while modeling user profiles, which demonstrates that the historical interaction plays a significant role in modeling user preference.
    
\end{itemize}

\subsection{Compatibility Matching} \label{sec:compatibilityModeling}

\subsubsection{\textbf{Evaluation Protocols}}
For each outfit in test dataset, we randomly select an item as the blank, and set three negative candidates. The target is to select the correct answer from four candidates to fill in the blank masked in the outfit. We report the accuracy to assess the performance.

\begin{table}
\centering
\caption{Overall performance comparison on compatibility matching.$*$ denotes the statistical significance for $p<0.001$.}
\label{tab:compatibilityCompare}
\resizebox{0.40\columnwidth}{!}{
\begin{tabular}{l|c }
\hline
       & FLTB   \\
\hline
\hline
Random       & 0.2425  \\
SiameseNet   & 0.5039 \\
Bi-LSTM       & 0.6384  \\
FHN          & 0.7422 \\
NGNN         & $\Mat{0.8422}$ \\ 
\hline
HFGN   & ${\Mat{0.8797}^{*}}$  \\
\hline
\end{tabular}}
\vspace{-5px}
\end{table}

\subsubsection{ \textbf{Baselines}}
We compare our model with the following baselines:
\begin{itemize}
    \item \textbf{Random}: The result for FLTB is randomly selected from 4 candiates.  
    \item \textbf{SiameseNet}~\cite{siamese}: SiameseNet sends the a pair of items into the a siamese network to project them into a style space and compare the distance between them. The compatibility score of the whole outfit is calculated by averaging the pairwise similarities in our experiment. 
    \item \textbf{Bi-LSTM}~\cite{bi-lstm}: Bi-LSTM takes an outfit as a sequence. It applies a bidirectional LSTM to learn the compatibility and predict the masked item. 
    \item \textbf{FHN}~\cite{FHN}: FHN encodes the visual features with category encoders and then learn binary codes for item embeddings. 
    FHN only classifies the fashion items into $3$ categories (\ie up, bottom and shoes). 
    The outfit score is the mean of pairwise compatibility score of items within the outfit. 
    \item \textbf{NGNN}~\cite{NGNN}: NGNN exploits the category information to represent items as the nodes in the category graph. 
    Thereafter, the node embeddings are updated through a one-layer graph convolution and a GRU cell.
    Finally, NGNN calculates the scores from each node embedding and introduces a self-attention mechanism~\cite{self-attention} to add them together as the output.
    
\end{itemize}

\subsubsection{ \textbf{Performance Comparision}}
We report the accuracy of test results in Table~\ref{tab:compatibilityCompare}. 
From Table~\ref{tab:compatibilityCompare}, we have the following observations:

\begin{figure}[t]
\centering
\vspace{10px}
\includegraphics[width=0.95\columnwidth]{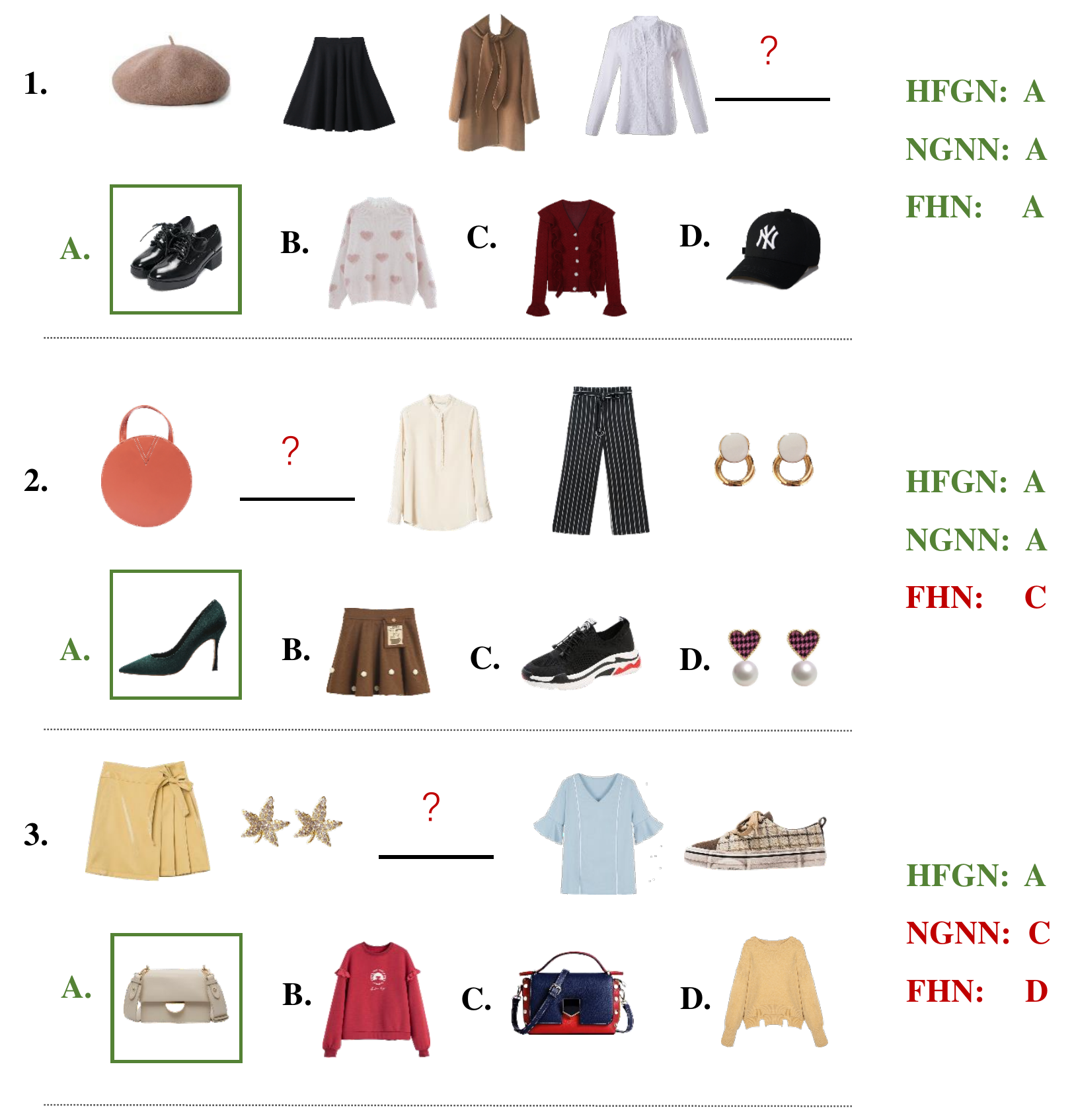}
\caption{Real examples of HFGN and two strong baselines (\ie NGNN and FHN) on Fill-in-the-Blank task. The green box circled the correct answer.}
\label{fig:caseStudy}
\end{figure}

\begin{itemize}
    \item Compared to other methods, SiameseNet achieves poor performance, indicating that only averaging the compatibility scores of item pairs is insufficient to learn high-order compatibility knowledge due to the overlook of the integrity of outfit, further limiting the performance. 
    
    \item We can see that Bi-LSTM performs better than SiameseNet. The reason might be that introducing Bi-LSTM can better learn the potential knowledge on compatibility. Compared to SiameseNet which directly averages the pairwise scores of items, Bi-LSTM takes the whole outfit as a sequence and learns high-order relations beyond item-level comparison. 
    
    \item FHN performs much better than SiameseNet, although it just averages the pairwise compatibility as well. Such improvement might be attributed to the introduced category knowledge. It verifies the importance of category information in modeling compatibility.
    Nevertheless, it only considers $3$ category labels, leading to miss compatibility relations on fine-grained categories.
    
    \item The two graph based methods, NGNN and HFGN, achieves better performance compared to other methods, indicating that the graph structure can better model complex interactions among items, further effectively inferring compatibility information.
    The improvement over Bi-LSTM and FHN method verifies the graph representation can better model the item interactions than sequence representation and pairwise representation.
    
    \item Our model, HFGN, achieves the best performance. Benefiting from the graph structure and message propagation across items, HFGN is capable of modeling complex interaction among items within the same outfit. 
    Compared to NGNN, we enhance item embedding with the compatibility information with its neighbours, the other items within the same outfit.
    Besides, we estimate the compatibility of outfits by introducing a R-view attention map, which can better capture the potential compatibility knowledge, further enhancing the model performance.   
\end{itemize}

\subsubsection{\textbf{Case Study}}
In figure~\ref{fig:caseStudy}, we visualize several test examples on Fill-in-the-Blank task to compare our model with the strong baselines (\ie NGNN and FHN).
From example 1, we can see that all the three models infer that the outfit lacks of a pair of shoes and correctly select the answer. From example 2, we can see that although FHN correctly selects complementary category (\ie shoes), it misses the correct answer as it just considers 3 coarse-grained category information and is incapable of exploiting the fine-grained category information (\eg the compatibility distinction between high-heels and sports shoes). 
From example 3, we can see that both NGNN and HFGN have inferred that the query outfit lack of a bag. Nevertheless, HFGN chooses a more compatible bag than NGNN.
That might be attributed to the more expressive propagation rules and compatibility formulation in HFGN.
These examples demonstrate the rationality and effectiveness of our model on compatibility matching task.

\section{Related Work}
In this section, we first review works on graph neural networks, and then we introduce the two related tasks in our work: personalized recommendation and compatibility matching.

\subsection{Graph Neural Network}
Graph neural networks (GNNs) are a kind of structure to model a set of elements and their relations.
Due to its great expressive power, GNNs have been widely used in many tasks involving rich relations including molecular analysis~\cite{MPNN,Protein}, 
image retrieval~\cite{Richang01,Richang02}
and visual comprehension~\cite{CMAT}.

Gori~\etal~ are the first to propose graph neural networks (GNNs)~\cite{GNN} which are capable of directly processing graphs to retain topological information.
Although GNNs can be applied on most types of graphs, \eg acyclic, cyclic, directed and undirected, 
the primitive GNNs have difficult to train for a fixed point.
Thereafter, GCN~\cite{GCN} was proposed to generalize convolutions to graph domain, which has achieved great success.
GCN can perform a convolution on the graph and aggregate the information derived from all the neighbours to update the node embeddings.
While distinct from GCN, GraphSAGE~\cite{graphsage} updates a node embedding by uniformly sampling and aggregating features from its local neighbours.
Although GNNs can update a node embedding by propagating information from neighbours with arbitrary depth,
a long-term message propagation might cause some problems.
To remedy this, recent advanced works~\cite{GGNN} attempt to introduce 
Gate Recurrent Units (GRU) in the propagation process. 

Most of the methods mentioned above have been proposed to solve node-focused problems. However, the edges in the graph may also contain rich information. In order to model relations between nodes, 
TransE~\cite{TransE} has been proposed to embed a graph into a continuous space, where each entity is represented as a point and each relation is modeled as a translating operation. However, TranE has flaws in dealing with complex relations, \eg 1-to-N, N-to-1 and N-to-N. To overcome this flaw, Wang \etal~ proposed TransH~\cite{TransH} to model a relation as a translating operation on a hyperplane. 

Because of the powerful modeling capabilities for complex relationships, GNNs have been widely used in personalized recommendation and fashion analysis, which will be introduced in the following.

\subsection{Personalized Recommendation}
Recommender system has been widely deployed to capture user preference in online platforms.
There have been a series of works committed to model user behavior effectively~\cite{EAR,MF,NCF,NGCF}.

MF~\cite{MF,BPRMF} is one of the most popular techniques, which maps each user and item as a vector with ID information and models the user-item interaction with inner product.

In order to enhance the model performance, some works attempted to incorporate side information into the prediction model.
For instance, VBPR \cite{VBPR} incorporated visual features of products to enhance item representations. 
NSCR~\cite{NSCR} utilized social relations to help model user behaviors.
Besides, a recent work~\cite{KPRN} introduced knowledge graph to provide complementary information while making recommendations.
Apart from the works mentioned above, some effort devoted to model user-item interaction by exploiting deep learning techniques. As inner product only linearly models the interaction behavior between users and items, they are insufficient to explore nonlinear and complicated relationships. To remedy this, He \etal~proposed NCF~\cite{NCF} to capture nonlinear relationships by leveraging a multi-layer perceptron.

Due to the great expressive power of graphs on modeling complex relations, some graph based recommender systems have been proposed recently. 
For instance, an advanced model, HOP-Rec~\cite{Hoprec}, enriched the training data by performing random walks on the interaction graph. 
NGCF~\cite{NGCF} stacked multiple propagation layers to aggregate high-order information into node representations. 
KGAT~\cite{KGAT} integrates user-item interaction graph and knowledge graph by linking items with their attributes. In such a hybrid architecture, node embeddings can be recursively refined by message propagation.
In addition, one recent research, LightGCN~\cite{LightGCN}, has found that two common designs in GCNs---feature transformation and nonlinear activation--- might degrade the recommendation performance and discarding these operations would benefit the model effectiveness.

\subsection{Compatibility Matching}
Fashion analysis~\cite{DeepFashion2,NGNN,MVC,LiaoMM18} has attracted increasing attention in recent years. In this paper, we focus on one fashion related task, compatibility matching.

Some works focus on estimating pairwise compatibility between a pair of items. McAuley \etal~proposed to map the items into a style space where compatible items will fall close to each other~\cite{Stylespace}. Following that, Veit \etal~fed a pair of item images into an end-to-end siamese network and measured the distance between them.
He \etal~proposed Monomer~\cite{Monomer} to model heterogeneous relationships of items beyond mere visual similarity.
Except for visual features, recent works highlight the importance of exploiting multi-modality features in fashion related task. 
For instance, Song \etal~utilized multiple modalities (\eg visual and contextual modalities) to learn a latent compatibility space~\cite{NeuroStylist} via a dual autoencoder network.
Afterwards, they proposed a knowledge distillation scheme~\cite{AKD} to learn from both data samples and domain knowledge so that the general knowledge rules can play a role as a teacher to guide the training process.

Recently, outfit related tasks have aroused interest in fashion community and there has been much effort devoted to model the compatibility of the whole outfit.
The key of modeling outfit compatibility is to represent an outfit properly. 
Han \etal~represented an outfit as a sequence and exploited bidirectional LSTMs to predict the compatibility score of one outfit~\cite{bi-lstm}. Benefiting from the BiLSTM architecture, the model can also predict a compatible item to fill the outfit with a blank.
There are also some works represent an outfit as a graph~\cite{TransNFCM,NGNN}. 
In NGNN~\cite{NGNN}, Cui \etal~represented an outfit as a graph to model complex relations among items, which has been demonstrated to be more effective than sequence and pairwise representation.

\section{Conclusion}
In this paper, we have proposed a new framework, Hierarchical Fashion Neural Network (HFGN), to solve the task of personalized outfit recommendation, which requires the recommended targets not only have a nice compatibility but also meet user's personal taste.
We build a hierarchical graph structure upon user-outfit interactions and outfit-item mappings.
The graph consists of three levels (\ie user, outfit and item level), where each level contains the corresponding type of nodes.
Through the message propagation on such hierarchical graph, the representations of nodes on each level can be refined by aggregating the interaction information derived from their neighbours.
By introducing ID embedding for the outfit, we incorporate outfit-level semantic into the outfit representation, which has been overlooked by the previous works~\cite{FPITF,FHN}.
And distinct from these works which separately considers compatibility matching and personalized recommendation, we regard the compatibility information as a passing message in the graph and encode it into the node representation. 
Extensive experiments have demonstrated the rationality and effectiveness of HFGN.

This work explores the potential of graph neural network in personalized outfit recommendation task. 
Benefiting from such hierarchical graph and message propagation, we can extend our work by incorporating multiple features (\eg textual feature) into the graph network to refine the node embeddings and model higher-order relations among them. 
In future work, we will focus on exploring the relationship between the two tasks (\ie outfit recommendation and compatibility modeling) and devising a more effective formulation to exploit this knowledge.
\begin{acks}
This work was supported by the National Key Research and Development Project of China (No. 2018AAA0101900), 
the National Natural Science Foundation of China (U19B2043, 61976185, U19A2079), 
Zhejiang Natural Science Foundation (LR19F020002, LZ17F020001), 
the Fundamental Research Funds for the Central Universities, 
and Chinese Knowledge Center for Engineering Sciences and Technology. 
Long Chen was supported by 2018 Zhejiang University Academic Award for Outstanding Doctoral Candidates.
This research is also part of NExT++ project, supported by National Research Foundation, Prime Minister's Office, Singapore under its IRC@Singapore Funding Initiative.
\end{acks}

\bibliographystyle{ACM-Reference-Format}
\bibliography{myrefer}

\end{document}